\documentclass{aa}  

\usepackage{graphicx}
\usepackage[varg]{txfonts}
\usepackage{subfig}
\usepackage{newtxtext,newtxmath}
\usepackage[labelfont=bf]{caption}

\usepackage{longtable,booktabs}

\usepackage{pdfpages}

\newcommand\R{\mathcal{R}}
\newcommand\Mstarp{M_\star^{\rm p}}
\newcommand\Mstar{M_\star}
\newcommand\Mstarstar{M_\star^*}
\newcommand\Mstarini{M_\star^{\rm initial}}
\newcommand\Mstaraccr{M_\star^{\rm accr}}
\newcommand\Mstarfin{M_\star^{\rm final}}
\newcommand\Mstarzzero{M_{\star,z=0}}
\newcommand\Mstarzthree{M_{\star,z=3}}
\newcommand\Rezzero{R_{{\rm e},z=0}}
\newcommand\Reobs{R_{\rm e,obs}}
\newcommand\sigmaeobs{\sigma_{\rm e,obs}}
\newcommand\Remajmin{R_{\rm e,maj+min}}
\newcommand\Mstarmajmin{M_{\rm \star,maj+min}}
\newcommand\sigmaemajmin{\sigma_{\rm e,maj+min}}
\newcommand\Rezthree{R_{{\rm e},z=3}}
\newcommand\sigmaezzero{\sigma_{{\rm e},z=0}}
\newcommand\sigmaezthree{\sigma_{{\rm e},z=3}}
\newcommand\Mzero{M_0}
\newcommand\Mmajstar{M_{\rm maj,\star}}
\newcommand\Mminstar{M_{\rm min,\star}}
\newcommand\fmaj{f_{\rm maj}}
\newcommand\fmin{f_{\rm min}}
\newcommand\fmini{f_{\rm mini}}
\newcommand\fenv{f_{\rm env}}
\newcommand\ximaj{\xi_{\rm maj}}
\newcommand\ximin{\xi_{\rm min}}
\newcommand\sigmavir{\sigma_{\rm vir}}
\newcommand\Mbh{M_{\bullet}}

\newcommand\sigmae{\sigma_{\rm e}}
\newcommand\sigmaeini{\sigma_{\rm e}^{\rm initial}}
\newcommand\sigmaefin{\sigma_{\rm e}^{\rm final}}

\newcommand\sigmar{\sigma_r}
\newcommand\sigmazero{\sigma_0}
\newcommand\sigmarsq{\sigma_r^2}

\newcommand\sigmalossq{\sigma_{\rm los}^2}
\newcommand\sigmaasq{\sigma_{\rm a}^2}
\newcommand\sigmaa{\sigma_{\rm a}}

\newcommand\Sigmastar{\Sigma_\star}
\newcommand\Sigmastarzero{\Sigma_{\star,0}}

\newcommand\rhotot{\rho_{\rm tot}}

\newcommand\rs{r_{\rm s}}
\newcommand\Mdm{M_{\rm dm}}
\newcommand\fdme{f_{\rm dm,e}}
\newcommand\rhodm{\rho_{\rm dm}}

\newcommand\rhos{\rho_{\rm s}}
\newcommand\Sersic{{S\'ersic }}
\renewcommand\Re{R_{\rm e}}
\newcommand\Reini{R_{\rm e}^{\rm initial}}
\newcommand\Refin{R_{\rm e}^{\rm final}}
\newcommand\Reaccr{R_{\rm e}^{\rm accr}}
\newcommand\Xiaccr{\Xi^{\rm accr}}
\newcommand\rg{r_{\rm g}}
\newcommand\vu{v_{\rm u}}

\newcommand{\kms}{\,{\rm km\,s^{-1}}}
\newcommand{\kpc}{\,{\rm kpc}}
\newcommand{\Msun}{{\rm M}_\odot}
\newcommand\Sect{Sect.\ }
\newcommand\Sects{Sects.\ }
\newcommand\Eq{Eq.\ }
\newcommand\Eqs{Eqs.\ }
\newcommand\Fig{Fig.\ }

\newcommand\rhostar{\rho_\star}
\newcommand\rhostarini{\rho_\star^{\rm initial}}
\newcommand\rhostareini{\rho_{\rm \star,e}^{\rm initial}}
\newcommand\rhostarfin{\rho_\star^{\rm final}}
\newcommand\rhostaraccr{\rho_\star^{\rm accr}}
\newcommand\rhodmini{\rho_{\rm dm}^{\rm initial}}
\newcommand\rhodmfin{\rho_{\rm dm}^{\rm final}}
\newcommand\rhodmaccr{\rho_{\rm dm}^{\rm accr}}

\newcommand{\beq}{\begin{equation}}
\newcommand{\eeq}{\end{equation}}

\newcommand\dd{{\rm d}}

\newcommand\alphasigma{{\alpha_\sigma}}
\newcommand\alphaR{{\alpha_R}}
\newcommand\alphaSMF{{\alpha_{\rm SMF}}}
\newcommand\alphabh{{\alpha_{\bullet}}}
\newcommand\betaR{\beta_{R}}

\usepackage[]{hyperref}
\hypersetup{colorlinks=true,linkcolor=red,citecolor=blue,urlcolor=magenta}

\defcitealias{Nip23}{N23}

\begin{document}

\titlerunning{Evolution of quiescent galaxies}
\title{Evolution of massive quiescent galaxies via envelope accretion}

\authorrunning{C.\ Nipoti}

\author{Carlo Nipoti}

\institute{
 Dipartimento di Fisica e Astronomia ``Augusto Righi'', Alma Mater Studiorum - Università di Bologna, via Gobetti 93/2, 40129, Bologna, Italy\\
\email{carlo.nipoti@unibo.it}\\
}

\date{April 2, 2025}

\abstract{}{Massive quiescent galaxies at high redshift are
  significantly more compact than their present-day counterparts. We
  investigate the roles, in determining this evolution, of major and
  minor mergers, and of the accretion of diffuse envelopes of stars and
  dark matter.}{We model the evolution in stellar mass ($\Mstar$),
  effective radius ($\Re$), and effective stellar velocity dispersion
  ($\sigmae)$ of a representative massive quiescent galaxy from
  $z\approx 3$ to $z\approx 0$, and compare the model with the
  observed redshift-dependent $\Re$-$\Mstar$ and $\sigmae$-$\Mstar$
  relations. In the model we account for the effects of collisionless
  (dry) major (satellite-to-main galaxy mass ratio $\xi>1/4$) and
  minor ($1/10<\xi<1/4$) mergers, using analytic recipes consistent
  with the results of $N$-body simulations of binary mergers. For the
  poorly constrained mini mergers ($\xi<1/10$) we explore both a
  'standard' model (based on the same assumptions used in the case
  of higher-$\xi$ mergers), and a heuristic 'envelope accretion'
  model, aimed at describing the case in which diffuse satellites are
  completely disrupted in the galaxy outskirts.}{Major and minor dry
  mergers, at rates estimated observationally from galaxy-pair counts,
  induce relatively small variations in $\Re$ and $\sigmae$, accounting
  only for $\approx 6\% $ of the size evolution and $\approx 40\%$ of
  the velocity-dispersion evolution observed from $z\approx 3$ to
  $z\approx 0$.  As an addition to major and minor dry mergers, envelope
  accretion performs better than standard mini mergers at reproducing
  the redshift-dependence of the $\Re$-$\Mstar$ and $\sigmae$-$\Mstar$
  relations, being also consistent with plausible evolutionary
  scenarios of scaling relations involving the mass of the central
  supermassive black hole.}{}

\keywords{Galaxies: elliptical and lenticular, cD -- galaxies: evolution -- galaxies: formation --  galaxies: kinematics and dynamics -- galaxies: structure}

   \maketitle

 \section{Introduction}
 \label{sec:intro}
 
The effective radius, $\Re$, and the central stellar velocity dispersion,
$\sigmazero$, of present-day massive quiescent galaxies (or early-type
galaxies; ETGs) correlate with the galaxy total stellar mass, $\Mstar$
\citep[][]{She03,Hyd09}.  These empirical scaling relations are found
to evolve with redshift, $z$, at least out to $z\approx 2-3$: the evolution
is such that, at given $\Mstar$, quiescent galaxies at higher redshift
have, on average, systematically smaller $\Re$
\citep[e.g.][]{van14,Mar24} and higher $\sigmazero$ \citep[e.g.][but
  see also \citealt{Dam22}]{van13,Can20}.  Recent observations with
the {\em James Webb Space Telescope} revealed that the size evolution
of quiescent galaxies occurs also at earlier times, being possibly
even stronger at redshift higher than 3
\citep{Ito24,Ji24,Wei24,Wri24}.

The main mechanism that is believed to be responsible for this
observed size and velocity-dispersion evolution is the growth of
individual galaxies via dissipationless (dry) merging, which has the
effect of increasing $\Re$ linearly or super-linearly with $\Mstar$,
while keeping constant or decreasing $\sigmazero$
\citep{Nip03,Naa09,Hil12,Hil13}.

The observed size evolution is however so strong that it is hard to
explain only with dry merging \citep{Nip09a}, also for merger histories with mass ratios and rates expected from
cosmological simulations \citep{Cim12,Nip12}.  Observational
estimates of the galaxy merger rates based on measurements of the
fraction of galaxy pairs are now available over wide redshift ranges
\citep{Con22,Dua24}. For $z\lesssim 3$, \citet{Con22}, comparing their
results with those of the Illustris cosmological simulation
\citep{Vol14}, find that the theoretically predicted
merger rates \citep{Rod15} overestimate the actual merger rates, at
least for mergers with mass ratio\footnote{We define the merger mass
ratio, $\xi\leq 1$, as the ratio between the less massive and the more
massive galaxy.}  $\xi>1/10$, even though the corresponding
theoretically predicted stellar mass accretion rate is lower than the
observational estimate \citep[see also][]{Mun17}.  This makes even more
puzzling the interpretation of the strong observed size evolution of
passive galaxies.

In this paper we compare the observed evolution of the $\Re$-$\Mstar$
and $\sigmazero$-$\Mstar$ relations of ETGs with the predictions of
dry-merging models, using observationally motivated merger mass growth
rates for major ($\xi>1/4$) and minor ($1/10<\xi<1/4$) mergers. We
then explore quantitatively the role of mini mergers ($\xi<1/10$), for
which we have recently started to have substantial observational
evidence \citep{Sue23}. The effect of mini mergers depends on the
poorly constrained properties of the involved satellites.  If they are
very diffuse, the cumulative effect of mini merger is expected to be an
'envelope accretion', that is the acquisition of of loosely bound stars
and dark matter (DM) in the galaxy outskirts \citep[as originally
  envisaged by][]{Ose10}.

Present-day ETGs lie also on scaling relations involving the mass of their
central supermassive black hole (BH), $\Mbh$, which is found to scale
linearly with $\Mstar$ \citep{Mag98} and with $\sigmazero$ as a power
law with index $4-5$ \citep{Fer00,Geb00}.  The evolution of these
scaling relations is poorly constrained observationally, because of
the difficulty of measuring BH masses in higher-$z$ ETGs. However, an
indication of the masses of central BHs in high-$z$ quiescent galaxies
comes from measurements of BH masses in the so-called relic galaxies
\citep{Tru14}, which are compact, massive present-day quiescent
galaxies with old stellar populations, believed to be descendants of
high-$z$ compact galaxies that evolved passively down to $z=0$ without
merging with other galaxies \citep[see also][]{Har25}. Relic galaxies
tend to have central BHs that are 'overmassive' (i.e.\, much more
massive than expected for their stellar mass), but in line with the
expectations for their high stellar velocity dispersion
\citep[e.g.][]{Com23}.  The hypothesis that high-$z$ ETGs host
overmassive BHs seems supported by the finding that, for galaxies
hosting active galactic nuclei, the ratio $\Mbh/\Mstar$ was higher at
$1\lesssim z \approx 3$ than today \citep{Mez24}. Instead, at least
with currently available observations, there is no evidence for
evolution of the $\Mbh$-$\sigmazero$ correlation \citep{She15}.  The
$\Mbh$-$\Mstar$ and $\Mbh$-$\sigmazero$ relations and their evolution represent
a complementary tool to test galaxy evolution models
\citep{Cio01,Nip03,Cio07}.  We thus compared the models discussed
in this work also including information on the central BH mass.

Finally, another evolutionary path of quiescent galaxies is that their
central DM fraction (for instance measured within $\Re$) tend to
increase as cosmic time goes on \citep{Men20,Dam22,Tor22}.  Any
successful model of the evolution of ETGs must account also for this
trend, which we thus took into account in our analysis.

This paper is organized as follows. In \Sect\ref{sec:obs} we report
the relevant observational data. In \Sect\ref{sec:effect} we introduce
a theoretical description of the expected effect of dry mergers on the
relevant galaxy properties. The predictions for mergers with mass
ratio $\xi>1/10$ are compared with observations in
\Sect\ref{sec:minmajcomparison}. Sections \ref{sec:mini} and
\ref{sec:envelo} consider the contribution of mini mergers ($\xi<1/10$).
The evolution of the BH mass and DM fraction is discussed in
\Sect\ref{sec:bhfdm}.  \Sect\ref{sec:concl} concludes.

\section{Observational data}
\label{sec:obs}

Here we describe the redshift-dependent observed scaling relations
that we considered in the present work, as well as the relevant
observational estimates of the merger mass growth rate for massive
galaxies. We limited ourselves to the redshift interval $0\lesssim
z\lesssim3$, over which the scaling relations are relatively well
characterized and over which \citet{Con22} provides information on the
observational estimates of the major and minor merger rates.

\subsection{Redshift-dependent size-stellar mass relation}
\label{sec:remstarz}

We took as reference for the observed size-stellar mass relation of
ETGs the work of \citet{van14}, who find that the redshift dependence of
the median correlation can be written as
\begin{equation}
 \log\left(\frac{\Re}{\kpc}\right)=\log A(z) +\beta(z)\log\left(\frac{\Mstar}{5\times 10^{10}\Msun}\right), 
\label{eq:size_mass}
\end{equation}  
where the dimensionless redshift-dependent coefficients $A$ and
$\beta$ are tabulated\footnote{The coefficient $\beta$ is called
$\alpha$ in \citet{van14} } in \citet{van14} for
$z=0.25,\,0.75,\,1.25,\,1.75,\,2.25,\,2.75$. While the logarithmic
slope, $\beta$, is weakly dependent on $z$ (with $0.7\lesssim
\beta\lesssim0.8$), the normalization, $A$, depends strongly on $z$ as
\begin{equation}
  A\propto(1+z)^{-1.48}.
  \label{eq:norm_size_mass}
\end{equation}
The above size-mass relation is plotted for different values of $z$ in
the upper panel of \Fig\ref{fig:scaling_laws_minmaj}, where the curves
for $z=0$ and $z=3$ were obtained by adopting, respectively, the same
values of $\beta$ as for $z=0.25$ and $z=2.75$, and extrapolating $A$
from, respectively, $z=0.25$ and $z=2.75$, using
\Eq(\ref{eq:norm_size_mass}). {The figure also shows the intrinsic
  scatter of the $z=0$ and $z=3$ relations, assumed to be the same as
  that reported by \citet{van14} at $z=0.25$ and $z=2.75$,
  respectively.}

{The redshift dependence of the $\Re$-$\Mstar$ relation of ETGs
  has been studied by various authors. For massive ETGs \cite{Dim19}
  and \citet{Mar24} find a variation with $z$ in $\Re$ at given
  $\Mstar$ consistent with that estimated by \citet{van14}.  A somewhat
  weaker evolution of $\Re$ is found by \citet{Mow19} and
  \citet{Ned21}.  \citet{Sue19} and \citet{Mil23} argue that, due to
  the presence of color gradients, the redshift dependence of the
  half-mass radius of massive quiescent galaxies is weaker than
  estimated by \citet{van14}, but this claim has been recently questioned
  by \citet{van24}. In any case, the studies that report weaker size
  evolution find, compared to \citet{van14}, at most $25\%$ smaller
  variation in $\Re$ at given $\Mstar$ over the redshift range
  $0\lesssim z\lesssim 3$. We account for this uncertainty when
  comparing models with observations in
  \Sect\ref{sec:minmajcomparison}.}

\subsection{Redshift-dependent velocity-dispersion stellar mass relation}
\label{sec:sigmamstarz}

When considering the central stellar velocity dispersion, it is
important to specify the aperture within which it is
measured. Hereafter, we indicate with $\sigmae$ the effective velocity
dispersion, that is the central velocity dispersion measured within an
aperture that can be approximated by a circular aperture of radius
$\Re$.

\begin{figure}
  \includegraphics[width=0.5\textwidth]{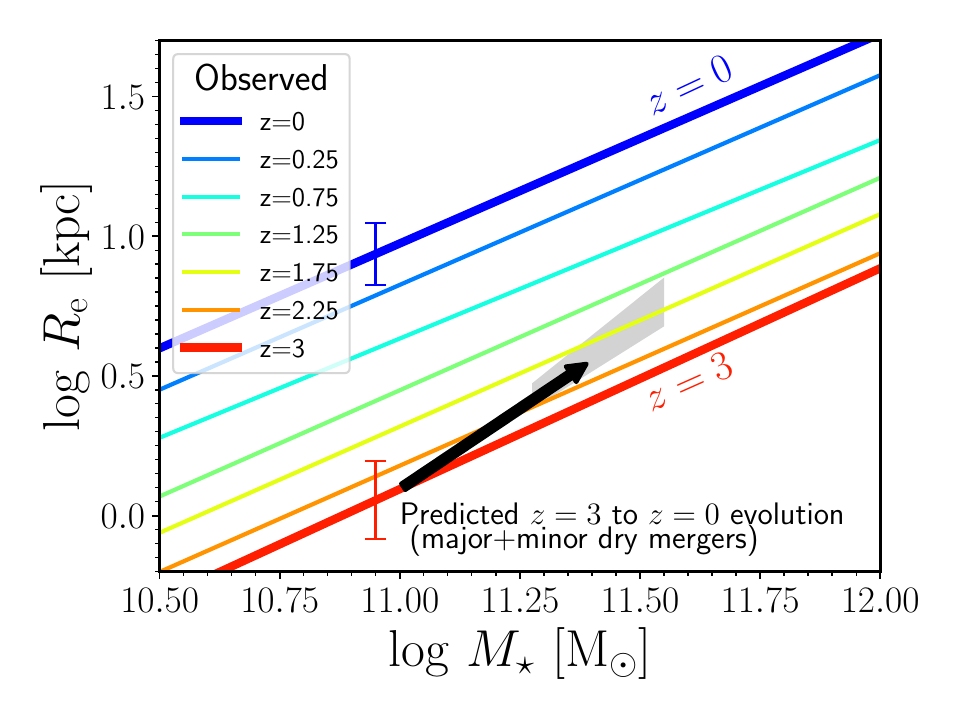}\\ \includegraphics[width=0.5\textwidth]{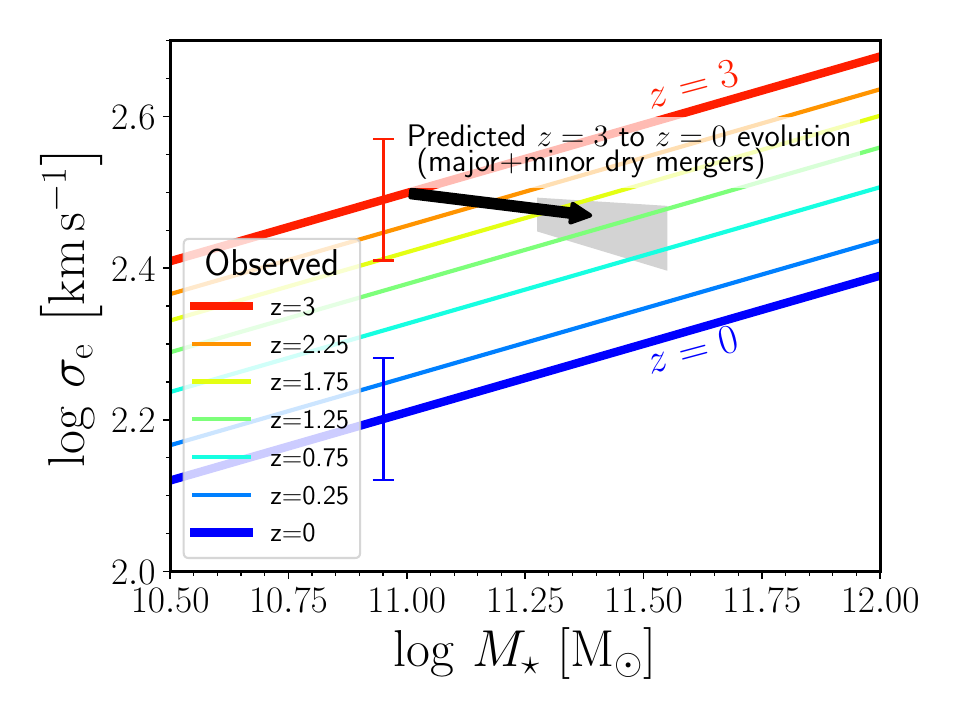}
  \caption{ETG evolution in stellar mass, effective radius, and
    effective velocity dispersion, due to major and minor
    mergers. Upper panel: Effective radius, $\Re$, as a function of
    stellar mass, $\Mstar$. The lines indicate the median observed
    $\Re$-$\Mstar$ relations of ETGs at different redshift
    (\citealt{van14}).  Lower panel: Effective stellar velocity
    dispersion, $\sigmae$, as a function of stellar mass, $\Mstar$. The
    lines indicate the median observed $\sigmae$-$\Mstar$ relations of
    ETGs at different redshifts (\citealt{Can20}). In each panel the
    black arrow indicates the predicted evolution of a representative
    model ETG from $z=3$ (tail of the arrow) to $z=0$ (head of the
    arrow), under the effect of major ($\xi>1/4$) and minor
    ($1/10<\xi<1/4$) dissipationless mergers occurring at the rate
    observationally estimated by \citet{Con22}. The grey area
    represents the uncertainty in the position of the head of the
    arrow. {The blue and red error bars indicate the intrinsic scatter
      of the scaling relations at $z=0$ and $z=3$, respectively.}}
    \label{fig:scaling_laws_minmaj}
\end{figure}

We took as reference for the observed redshift-dependent
$\sigmae$-$\Mstar$ relation of ETGs the results of \citet{Can20},  and
in particular their extended-sample, constant-slope non-evolving
scatter model, for which the median $\sigmae$ is given by
\begin{equation}
\log\left(\frac{\sigmae}{\kms}\right)\simeq 2.21+0.18 \log\left(\frac{\Mstar}{10^{11}\Msun}\right)+0.48\log(1+z).
\end{equation}
Though this fit was obtained for galaxies in the redshift range
$0\lesssim z \lesssim 2.5$, we extrapolated it out to $z=3$ when
plotting, in the lower panel of \Fig\ref{fig:scaling_laws_minmaj},
the median $\sigmae$-$\Mstar$ curves for the same selection of
redshifts as in the upper panel of the same figure.  {The figure
  also shows the intrinsic scatter in $\sigmae$ of the $z=0$ and $z=3$
  relations, assumed to be $0.08$ dex, the redshift-independent value
  found by \citet{Can20}.}

{As far as we know, \citet{Can20} is the only work in which the
  $\sigmae$-$\Mstar$ relation of massive ETGs is studied
  systematically as a function of redshift. However, different studies
  have investigated whether and how $\sigmae$ depends on $z$ for
  galaxies of similar $\Mstar$. A few authors, consistent
  with \citet{Can20}, find that $\sigmae$ at given $\Mstar$ is higher
  at higher $z$ \citep[][]{Cap09,Cen09,van13,Bel17,Sto20}, while
  others find no evidence of velocity-dispersion evolution
  \citep{Tan19,Dam22}.  We discuss also the hypothesis of
  redshift independent $\sigmae$-$\Mstar$ relation in our analysis in
  \Sect\ref{sec:minmajcomparison}.  }

\subsection{Observed major- and minor-merger mass growth rates}
\label{sec:mergerrate}

{\citet{Con22} measured the galaxy pair and merger fractions as
  functions of redshift and stellar mass ratio for a large sample of
  massive galaxies in the redshift range $0<z<3$.  Based on their
  merger rate estimates (which probe as minimum mass ratio
  $\xi=1/10$), they find that galaxies with stellar mass
  $\Mstar>10^{11}\Msun$, from $z=3$ to $z=0$, grow in stellar mass by
  $\fmaj=93^{+49}_{-31}\%$ via major mergers ($\xi>1/4$) and by
  $\fmin=29^{+17}_{-12}\%$ via minor mergers ($1/10<\xi<1/4$).} It is
worth noting that these estimates are based on a stellar-mass
selection; when \citet{Con22} apply instead a constant-number density
selection, they find significantly lower stellar-mass growth rates for
both major and minor mergers.  {In the spirit of maximizing the
  possible effect of mergers on the evolution of the scaling
  relations, in our analysis we assumed that, on average, from $z=3$
  to $z=0$ massive ETGs increase their stellar mass by the
  aforementioned values of $\fmaj$ and $\fmin$, for major and minor
  mergers, respectively.}  It is important to stress that the
contribution of mini mergers ($\xi<1/10$) is not included in these
numbers.

{Before \citet{Con22}, other authors attempted to measure the
  merger rate of massive galaxies at $z\lesssim 3$, accounting for the
  contribution of major and minor mergers.  \citet{Blu12} estimate
  that, from $z=3$ to the present (see also \citealt{Lop11}), galaxies
  of $\Mstar\sim10^{11}\Msun$, via mergers with mass ratio
  $\xi>1/100$, undergo a fractional stellar-mass increase $2.0\pm
  1.2$, which, within the relatively large uncertainty, is consistent
  with the estimate of \citet{Con22}. A somewhat lower stellar mass
  growth due to $\xi>1/10$ mergers ($\approx 50\%$ from $z=2.5$ to
  $z=0$) is instead estimated by \citet[]{Man16}.  Similar results are
  reported by \citet{New12}, who find merger rates consistent with
  those of \citet[]{Man16} in the redshift range $0.4 < z < 2.5$. }

\section{Effect of dry mergers on $\Re$ and $\sigmae$} 
\label{sec:effect}

The effects of mergers on the galactic structural and kinematic
properties depend in a complex way on the internal properties of
the interacting galaxies, on their mass ratio and on the
characteristics of their mutual orbits \citep[see section 8.9
  of][]{CFN19}.

Even limiting to the case of gas-poor spheroidal galaxies, the
variations in $\Re$ and $\sigmae$ produced by mergers depend not only
on the merger mass ratio, but also on the orbital parameters of the
encounter and on the internal distribution of stars and DM of the
interacting systems.  However, some general physical arguments can be
used to obtain approximate analytic predictions for the evolution of
$\Re$ and $\sigmae$ in dry merging.

\subsection{Analytic formulae}

Under the assumption that
$\Re\propto\rg$ and $\sigmae\propto\sigmavir$, where $\rg$ is the
gravitational radius and $\sigmavir$ is the virial velocity
dispersion, the effect of dry mergers on $\Re$ and $\sigmae$ can be
computed using energy conservation and the virial theorem
\citep{Nip03,Naa09,Hil12}. In particular, in the case of parabolic orbits,
under these hypotheses, for a galaxy growing with a sequence of dry
mergers with mass ratio $\xi$, $\Re$ and $\sigmae$ grow with stellar
mass as, respectively,
\begin{equation}
  \Re\propto\Mstar^{\alphaR(\xi)}
  \label{eq:remstar_dry}
\end{equation}
and
\begin{equation}
\sigmae\propto\Mstar^{\alphasigma(\xi)},
\label{eq:sigmstar_dry}
\end{equation}
with
\begin{equation} 
\alphaR(\xi)=2-\frac{\ln (1+\xi^{2-\betaR})}{\ln(1+\xi)}
\label{eq:alphaR}
\end{equation}
and
\begin{equation} 
\alphasigma(\xi)=-\frac{1}{2}\left[1-\frac{\ln (1+\xi^{2-\betaR})}{\ln(1+\xi)}\right]
\label{eq:alphasigma}
\end{equation} 
\citep{Nip12}, where we have assumed that the merging galaxies lie on
a size-mass relation $\Re\propto \Mstar^{\betaR}$.

\subsection{Range of validity of the analytic formulae for major, minor, and mini mergers}
\label{sec:validity}

There are a few works in the literature in which the evolution of the
effective radius and of the central velocity dispersion in
dissipationless mergers has been studied with $N$-body simulations
\citep{Nip03,Boy06,Nip09b,Hil12,Nip12,Hil13,Fri17,Ran24}.  {The
  simulations run in these works differ widely in mass ratio, orbital
  parameters, and internal properties of the interacting systems (such
  as amount and distribution of dark and baryonic matter), and
  consequently also the behaviour of $\Re$ and $\sigmae$ shows a
  relatively wide variety.  However, trying to make a synthesis of the
  results of these work,
  \Eqs(\ref{eq:remstar_dry}-\ref{eq:alphasigma}) appear to describe
  sufficiently well the average effects of major mergers
  \citep[][]{Nip03,Boy06,Nip09b,Hil12} and of minor mergers
  (\citealt{Nip12}; \citealt{Fri17}).}  In $N$-body simulations of
equal-mass mergers the central velocity dispersion tends to increase
slightly, but the deviation from the constant-$\sigmae$ prediction of
\Eq(\ref{eq:alphasigma}) is typically within 15\%
\citep{Nip03,Boy06,Nip09b,Hil12}.  {\citet{Ran24} find that the
  size evolution is stronger when central BHs are included in the
  simulations: on average, in their numerical experiments, the index
  $\alphaR$ is higher in the presence than in the absence of central
  BHs by $\approx 10\%$ in major mergers and $\approx 30\%$ in minor
  mergers (see their table 5).  Considering overall the results of the
  aforementioned works, we decided to use
  \Eqs(\ref{eq:remstar_dry}-\ref{eq:alphasigma}) for major and minor
  mergers, but keeping in mind the uncertainties on $\alphaR$ and
  $\alphasigma$.}

The situation is much more uncertain for mini mergers ($\xi<1/10$),
which have been studied relatively little in the literature.  As far
as we know, the only systematic exploration of the effect of mini
mergers on $\Re$ and $\sigmae$ is that of \citet{Hil12}, who simulated
dry mergers with mass ratios $0.05<\xi\leq 0.1$ in which the satellite
is much less dense than the main galaxy ($0\lesssim \betaR \lesssim
0.3$; see also \citealt{Hil13} and \citealt{Ran24}).  \citet{Hil12}
found significant deviations from the analytic predictions of
\Eqs(\ref{eq:remstar_dry}-\ref{eq:alphasigma}): in particular the
values of $\alphasigma$ found in simulations are systematically higher
than those predicted by \Eq(\ref{eq:alphasigma}).  Though limited to
some very specific cases, the results of these numerical experiments
suggest that \Eqs(\ref{eq:remstar_dry}-\ref{eq:alphasigma}) do not
necessarily capture the effects of mergers with very small mass
ratios.  Thus, in the attempt of bracketing the realistic behaviour,
in the case of mini mergers we explored two very different scenarios:
'standard mini mergers' (\Sect\ref{sec:mini}), in which we assumed
that \Eqs(\ref{eq:remstar_dry}-\ref{eq:alphasigma}) hold, and
'envelope accretion' (\Sect\ref{sec:envelo}), in which we assumed that
the satellites are very diffuse and they deposit the great majority of
their stars and DM in the outskirts of the main galaxy.

\section{Weak evolution driven by major and minor mergers at the observed rates}
\label{sec:minmajcomparison}

Using the formulae presented in \Sect\ref{sec:effect}, we computed the
evolution of individual galaxies in the $\Mstar$-$\Re$ and
$\Mstar$-$\sigmae$ planes, due to $\xi>1/10$ mergers at the observed
rate (\Sect\ref{sec:mergerrate}), and compared it with the observed
evolution of the scaling relations (\Sects\ref{sec:remstarz} and
\ref{sec:sigmamstarz}). { Given that the considered scaling relations
  are power laws with virtually redshift-independent
  slopes\footnote{While the slope of the considered $\sigmae$-$\Mstar$
  relation is strictly independent of $z$ \citep{Can20}, the slope of
  the $\Re$-$\Mstar$ relation of \citet{van14} varies only very slight
  with $z$ (see their table 1).}, and that the merger-driven evolution
  in the $\Mstar$-$\Re$ and $\Mstar$-$\sigmae$ planes is modelled with
  power laws, the initial stellar mass of our reference galaxy can be
  chosen arbitrarily in the mass range in which the scaling relations
  have been computed.}

However, to consistently use the merger mass growth
rates of \citet{Con22}, we limited ourselves to stellar masses
$\Mstar\geq 10^{11}\Msun$.  We thus considered as our 'initial galaxy'
a model ETG that at $z=3$ has $\Mstar=10^{11}\Msun$, $\Re\simeq
1.24\kpc$ and $\sigmae\simeq316\kms$, values such that it lies on the
$z=3$ scaling relations shown in \Fig\ref{fig:scaling_laws_minmaj}.
{We then modelled its evolution assuming that its stellar mass from $z=3$
to $z=0$ grows as
\begin{equation}
\Mstarzzero=\Mstarzthree(1+\fmaj)(1+\fmin),
\end{equation} 
where $\fmaj$ and $\fmin$ are the fractional stellar mass increase
\footnote{These quantities can be to a good approximation
identified with the ratios $\delta \Mmajstar /\Mzero$ and $\delta
\Mminstar /\Mzero$ computed by \citet{Con22}, because in that paper
$\Mzero$ is the average mass ($\Mstar\approx1.6\times 10^{11}\Msun$)
of the massive galaxy population in the redshift interval $0<z<3$.}
over the redshift range $0<z<3$, and the subscripts $z=0$ and $z=3$
indicate at which redshift the quantity is evaluated.}  It follows from
the equations of \Sect\ref{sec:effect} that the corresponding
evolution in size and velocity dispersion is described by
\begin{equation}
\Rezzero=\Rezthree(1+\fmaj)^{\alphaR(\ximaj)}(1+\fmin)^{\alphaR(\ximin)},  
\end{equation}
and
\begin{equation}
  \sigmaezzero=\sigmaezthree(1+\fmaj)^{\alphasigma(\ximaj)}(1+\fmin)^{\alphasigma(\ximin)},  
\end{equation}
respectively.
Here $\ximaj$ and $\ximin$ are characteristic mass ratios for major
and minor mergers, respectively, and we recall that $\alphaR$ and
$\alphasigma$ depend also on $\betaR$.

For our fiducial model we assumed $\fmaj=0.93$, $\fmin=0.29$
(\Sect\ref{sec:mergerrate}), $\betaR=0.76$ (the median value of the
$\Re$-$\Mstar$ relation logarithmic slope, $\beta$, found by
\citealt{van14}; \Sect\ref{sec:remstarz}), $\ximaj=(1+0.25)/2=0.625$
and $\ximin=(0.1+0.25)/2=0.175$.  With these parameters we obtained
$\delta\log\Mstar=0.40$, $\delta\log\Re=0.46$ and
$\delta\log\sigmae=-0.03$, where we have introduced the notation
$\delta q=q(z=0)-q(z=3)$, to indicate the variation in a generic
quantity $q$ for an individual galaxy.  In
\Fig\ref{fig:scaling_laws_minmaj} the black arrows indicate this
predicted evolution for our fiducial galaxy in the $\Mstar$-$\Re$
plane (upper panel) and $\Mstar$-$\sigmae$ plane (lower panel), with
the head of the arrow indicating the predicted position at $z=0$ due
to major and minor mergers only. In each panel, the grey shaded area
indicates the uncertainty in the position of the head of the arrow.
We computed this area considering the uncertainties on $\fmaj$ and
$\fmin$ (\Sect\ref{sec:mergerrate}), as well as those on $\alphaR$ and
$\alphasigma$, which we approximately accounted for by allowing $\ximaj$
and $\ximin$ to vary, respectively, in the ranges $1/4\leq\ximaj\leq
1$ and $1/10\leq\ximin\leq 1/4$, and $\betaR$ in the range
$0.7\leq\betaR\leq0.8$ (\Sect\ref{sec:remstarz}).

{The result shown by the upper panel of
  \Fig\ref{fig:scaling_laws_minmaj} is striking: major and minor
  mergers at the observed rate are expected to produce a weak size
  evolution from $z=3$ to $z=0$, much less than the observed evolution
  of the size-mass relation over the same redshift interval (even if
  the intrinsic scatter of the relations and the uncertainty in the
  position of the head of the arrow are taken into account)}.  Our
representative model galaxy at redshift $z=0$ (head of the arrow in
the plot) has stellar mass $\simeq 2.5\times 10^{11}\Msun$ and
effective radius $\Re\simeq 3.6\kpc$, about a factor of five smaller
than the average $\Re$ of observed $z=0$ ETGs of similar stellar mass.
{This finding confirms and strengthens the results of
  \citet{New12} and \citet{Man16}, who reported that the observed
  merger rates are insufficient to explain the strong size evolution of
  massive quiescent galaxies.

  The lower panel of \Fig\ref{fig:scaling_laws_minmaj} shows that also
  the predicted evolution of the velocity dispersion is too weak: the
  $\sigmae$ of the model galaxy decreases slowly, with a final value
  as high as $\simeq294\kms$, about 50\% higher than the median
  $\sigmae$ of observed $z=0$ ETGs of similar stellar mass. However,
  different from the size evolution, the predicted velocity-dispersion
  evolution is marginally consistent with the observations if the
  intrinsic scatter of the relations and the uncertainty in the
  position of the head of the arrow are taken into account.  }

To quantify the relative contribution of major and minor mergers to the size
and velocity-dispersion evolution from $z=3$ to $z=0$, it is
convenient to define the offsets
\begin{equation}
\Delta\Reobs=|\Reobs(\Mstarmajmin,0)-\Reobs(\Mstarmajmin,3)|,
\end{equation}
\begin{equation}
\Delta\Remajmin=|\Remajmin-\Reobs(\Mstarmajmin,3)|,
\end{equation}
\begin{equation}
\Delta\sigmaeobs=|\sigmaeobs(\Mstarmajmin,0)-\sigmaeobs(\Mstarmajmin,3)|,
\end{equation}
and
\begin{equation}
\Delta\sigmaemajmin=|\sigmaemajmin-\sigmaeobs(\Mstarmajmin,3)|,
\end{equation}  
where $\Reobs(\Mstar,z)$ and $\sigmaeobs(\Mstar,z)$ are, respectively,
the observed $\Re$-$\Mstar$ and $\Re$-$\Mstar$ correlations
(\Sect\ref{sec:remstarz} and \ref{sec:sigmamstarz}), and
$\Mstarmajmin$, $\Remajmin$, and $\sigmaemajmin$ are the values of
$\Mstar$, $\Re$, and $\sigmae$ predicted at $z=0$ by a model accounting
only for major and minor mergers (corresponding to the heads of the
arrows in \Fig\ref{fig:scaling_laws_minmaj}). Over the range $0< z<
3$, the contribution of major and minor mergers is
$\Delta\Remajmin/\Delta\Reobs\simeq 6\%$ to the size evolution and
$\Delta\sigmaemajmin/\Delta\sigmaeobs\simeq 42\%$ to the
velocity-dispersion evolution.

{We have mentioned in \Sects\ref{sec:remstarz} and
  \ref{sec:sigmamstarz} that, according to some authors, the evolution
  in $\Re$ and $\sigmae$ at given $\Mstar$ could be weaker than
  estimated by \citet{van14} and \citet{Can20}. Even assuming that
  $\Reobs(\Mstarmajmin,0)$ is smaller by 25\% (see
  \Sect\ref{sec:remstarz}), $\Delta\Remajmin/\Delta\Reobs$ is as small
  as $\approx 8\%$. The evolution of the $\sigmae$-$\Mstar$ relation
  is much more uncertain and debated in the literature, so it is not
  excluded that $\Delta\sigmaemajmin/\Delta\sigmaeobs$ is closer to
  unity than found using the $z$-dependent scaling relation of
  \citet{Can20}, which is however the state of the art with currently
  available datasets.  The result that $\Delta\Remajmin/\Delta\Reobs$
  is small is robust also against uncertainties on the value
  $\alphaR$: even if, following \citet[][]{Ran24}, we increase
  $\alphaR$ by $10\%$ for major mergers and by $30\%$ for minor
  mergers (see \Sect\ref{sec:validity}), we get
  $\Delta\Remajmin/\Delta\Reobs \approx 11\%$.  }

In this section we have estimated the predicted evolution due only to
mergers with mass ratio $\xi>1/10$. To this evolution we must add the
contribution of mergers with $\xi<1/10$, which we consider in
\Sects\ref{sec:mini} and \ref{sec:envelo}.

\section{Standard mini dry mergers}
\label{sec:mini}

\begin{figure}
  \includegraphics[width=0.5\textwidth]{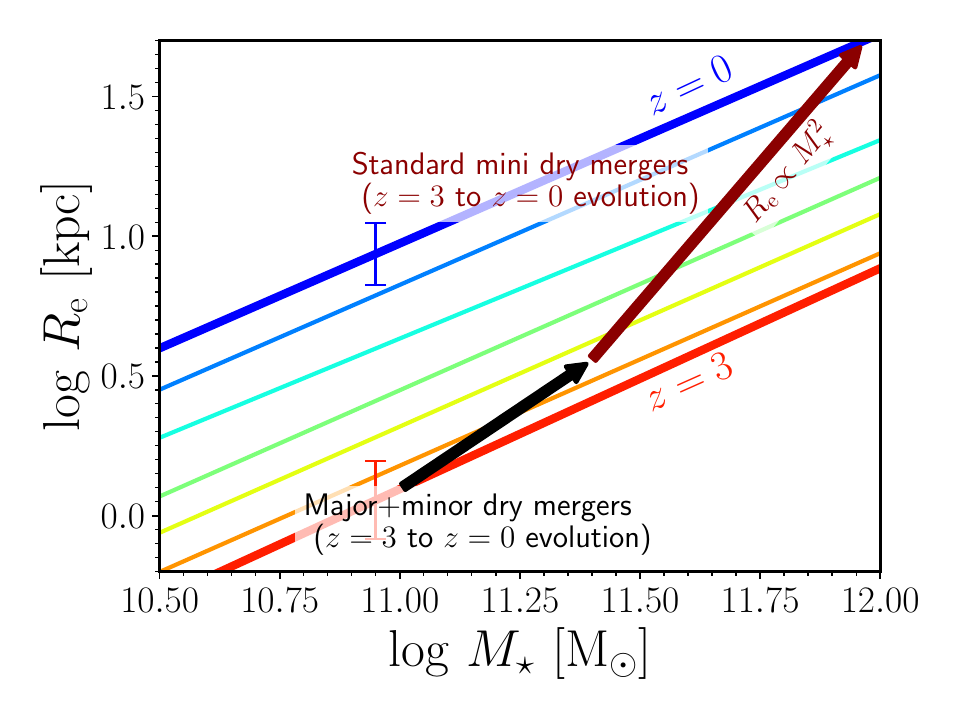}\\ \includegraphics[width=0.5\textwidth]{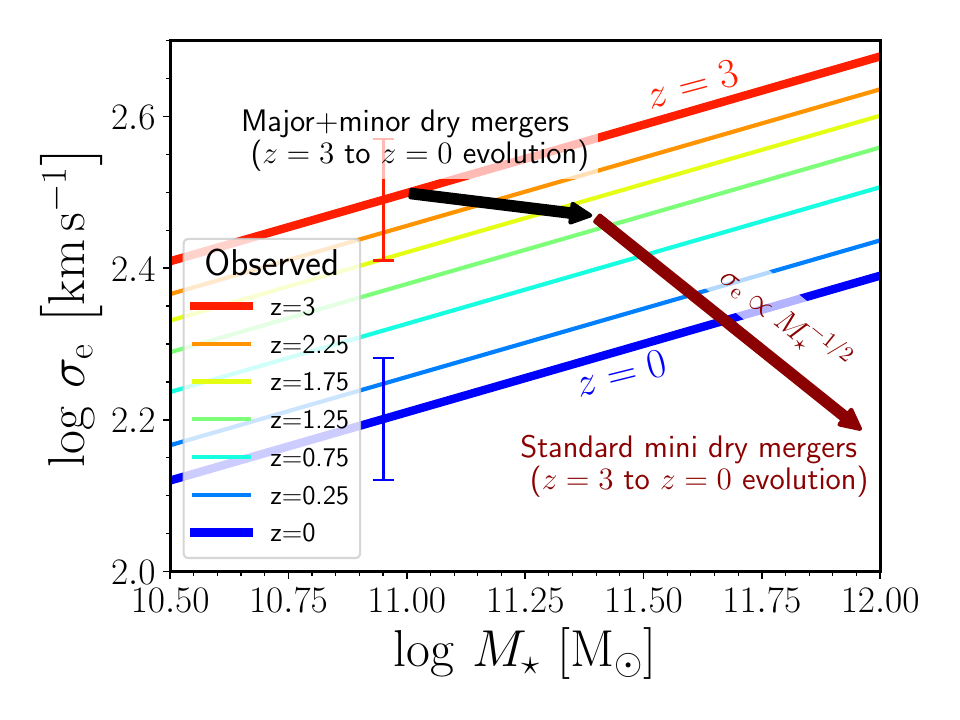}
  \caption{ETG evolution in stellar mass, effective radius, and
    effective velocity dispersion, due to major, minor, and standard
    mini mergers.  Same as \Fig\ref{fig:scaling_laws_minmaj}, but with
    the addition of the potential contribution from $z=3$ to $z=0$ of
    standard mini dry mergers (red arrows), that is mergers with
    $\xi\ll 1$ in which $\Re$ and $\sigmae$ vary with stellar mass as
    $\Re\propto\Mstar^2$ and $\sigmae\propto\Mstar^{-1/2}$,
    respectively. These power laws are the limits of
    \Eqs(\ref{eq:remstar_dry}) and (\ref{eq:sigmstar_dry}) for $\xi\to
    0$ (independent of $\betaR$, provided $\betaR<1$). While the
    stellar mass growth attributed to major and minor mergers (black
    arrows) is based on observational estimates
    (\Sect\ref{sec:mergerrate}), the stellar mass growth attributed to
    mini mergers (red arrows) is chosen ad hoc so that the head of the
    red arrow lies on the observed $z=0$ $\Re$-$\Mstar$ correlation
    (upper panel).}
    \label{fig:standard_mini}
\end{figure}

As pointed out in \Sect\ref{sec:effect}, the effect on $\Re$ and
$\sigmae$ of mass growth via mini ($\xi<1/10$) mergers is poorly
constrained. When considering growth via mini mergers we thus explored
two different simple scenarios. In this section we consider standard
mini mergers, while in \Sect\ref{sec:envelo} we consider envelope
accretion. With standard mini mergers we mean parabolic $\xi<1/10$ dry
mergers such that $\Re\propto\rg$ and $\sigmae\propto\sigmavir$, so
that the formulae of \Sect\ref{sec:effect} apply.  Given the
uncertainty on the dominant mass ratio of mini mergers and the fact
that the observed evolution of the $\Re$-$\Mstar$ relation is strong
(\Fig\ref{fig:scaling_laws_minmaj}, upper panel), to maximize the size
evolution we model here mini mergers as dry mergers with vanishingly
small mass ratio.  In the considered hypotheses, taking the limit for
$\xi\to 0$ of \Eqs(\ref{eq:remstar_dry}) and (\ref{eq:sigmstar_dry})
we get\footnote{In fact, this is true provided $\betaR<1$, which is
however a standard assumption, generally supported by the
observations.}  $\Re\propto\Mstar^2$ and $\sigmae\propto\Mstar^{-1/2}$
\citep{Naa09}. In \Fig\ref{fig:standard_mini} we show the evolution in
the $\Mstar$-$\Re$ (upper panel) and $\Mstar$-$\sigmae$ (lower panel)
planes of our representative model galaxy when, to the contribution of
major and minor mergers (\Sect\ref{sec:minmajcomparison}), we add a
contribution of standard mini mergers such that our model galaxy lies
on the median $\Re$-$\Mstar$ relation of ETGs at $z=0$. To achieve
this, our model galaxy grows in stellar mass by almost a factor of 10
from $z=3$ to $z=0$ (including major, minor, and mini mergers).
{The required fractional mass growth due to mini mergers is thus $\fmini\approx 2.6$.}
The
lower panel of \Fig\ref{fig:standard_mini} shows that the same model
galaxy is predicted to have way too low $\sigmae$ at $z=0$.

In \Fig\ref{fig:standard_mini} we show as example of evolution driven
by standard mini mergers only the case with vanishingly small mass
ratio ($\xi\to 0$), which maximizes the increase in $\Re$ and the
decrease in $\sigmae$ for given stellar mass growth.  If, instead of
$\xi\to 0$, we assumed for our standard mini-merger model as
characteristic mass ratio a more realistic finite value $\xi <1/10$,
we would get red arrows with shallower slope in both panels of
\Fig\ref{fig:standard_mini}. However, this would not help solve the
problem of the too low predicted $\sigmae$ at $z=0$, because higher
stellar mass growth via mini mergers would be required to match the
$z=0$ $\Re$-$\Mstar$ relation, and thus too low values of $\sigmae$ at
$z=0$ would be predicted even with a shallower predicted evolution in
the $\Mstar$-$\sigmae$ plane.

We recall that in \Sect\ref{sec:effect} (and thus in the standard mini
merger model here considered), we assumed not only that $\Re$ and
$\sigmae$ are proportional to the gravitational radius and virial
velocity dispersion, but also that the mergers are parabolic.  If the
orbital energy is negative, the final velocity dispersion can be
higher than in the parabolic case \citep{Nip09b,Pos14}.  However, the
more bound the orbits, the weaker is the size growth, so it seems
unlikely that adding the effects of negative orbital energy can help
produce, at the same time, the needed strong evolution in $\Re$ and
the relatively weak evolution of $\sigmae$, within the framework of
standard mini mergers.  A more promising solution is the accretion of
very diffuse satellites, which we discuss in the next section.

\section{Envelope accretion}
\label{sec:envelo}

We consider here the case in which in mini mergers the accreted
material end up in an envelope, that is a diffuse, extended component.

\subsection{A toy model for envelope accretion}
\label{sec:toy}

\begin{figure}
  \includegraphics[width=0.5\textwidth]{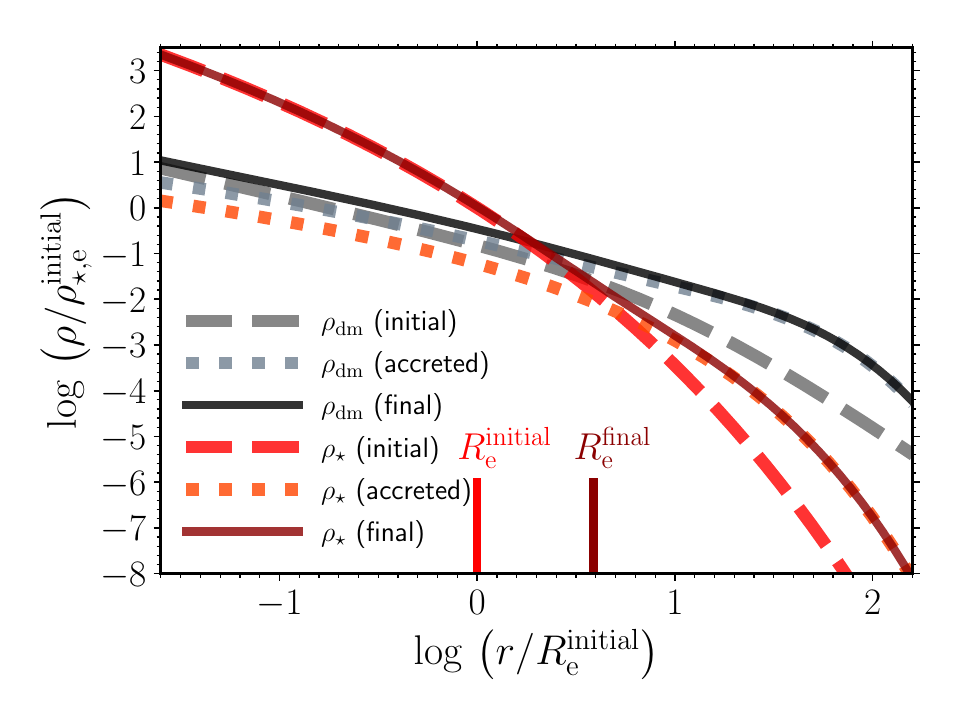}
  \caption{Envelope accretion-driven evolution of the density
    distribution of a representative ETG. Initial (dashed curves),
    accreted (dotted curves), and final (solid curves) stellar (red
    curves) and DM (black and grey curves) density profiles of our
    reference model galaxy growing via envelope accretion. The
    vertical lines indicate the initial ($\Reini$) and final
    ($\Refin$) effective radii. $\rhostareini$ is the initial stellar
    density at $r=\Reini$.}
    \label{fig:rho}
\end{figure}

We built a toy model for envelope accretion by considering an initial
galaxy model (representing a higher-$z$ quiescent galaxy) and a final
galaxy model (representing its lower-$z$ descendant) obtained by simply
adding to the initial galaxy model an accreted stellar component and
an accreted DM component. The models, which are spherically symmetric,
were computed following the approach of \cite{Nip08}. The initial
galaxy has stellar density $\rhostarini(r)$ and DM density
$\rhodmini(r)$; the final galaxy has stellar density
$\rhostarfin(r)=\rhostarini(r)+\rhostaraccr(r)$ and DM density
$\rhodmfin(r)=\rhodmini(r)+\rhodmaccr(r)$, where $\rhostaraccr$ and
$\rhodmaccr$ are, respectively, the stellar and DM accreted mass
density. We now define the functional forms that we use to model
$\rhostarini$, $\rhodmini$, $\rhostaraccr$, and $\rhodmaccr$.

\subsubsection{Stellar components}
\label{sec:stellarcomp}

Both the initial and the accreted stellar components are described by
\cite{Ser68} models. The projected density distribution follows the
\Sersic $R^{1/m}$ law
\begin{equation}
\Sigmastar(R)=\Sigmastarzero\,\exp\left[-b(m)\left(\frac{R}{\Re}\right)^{1/m}\right],
\end{equation}
where $b(m)\approx 2m-1/3+4/(405m)$ \citep{Cio99} and $m$ is the
so-called \Sersic index.  By deprojecting $\Sigmastar$ one obtains the
corresponding intrinsic density distribution \citep{Bin08}
\begin{equation}
\rhostar(r)=-{1\over\pi}\int_r^\infty{\dd\Sigmastar\over \dd R}{\dd R\over
    \sqrt{R^2-r^2}},
\end{equation}
which we computed numerically.

\subsubsection{Dark-matter components}
\label{sec:darkcomp}

The DM halo of the initial galaxy is
described by a \citet[][]{Nav96} model, with density
distribution 
\begin{equation}
  \rhodmini (r)=4\rhos\left(\frac{r}{\rs}\right)^{-1}\left(\frac{r}{\rs}+1\right)^{-2},
\end{equation}
where $\rs$ is the scale radius and $\rhos\equiv\rhodmini(\rs)$.

The dark accreted component is assumed to have a density distribution
\begin{equation}
\rhodmaccr(r)=\Xiaccr\rhostaraccr(r)\left(1+\frac{r}{\Reaccr}\right)^3,
\label{eq:rhodmaccr}
\end{equation}
where $\Reaccr$ is the projected half-mass radius of the accreted
stellar component and $\Xiaccr$ is a dimensionless factor such that,
for given $\rhostaraccr(r)$, the DM-to-stellar mass ratio of the
accreted material increases for increasing $\Xiaccr$.

\subsubsection{Stellar mass, effective radius, and central velocity dispersion}
\label{sec:observables}

For both the initial and final galaxy models we need the
total mass, the effective radius, and the effective stellar velocity
dispersion.  The total stellar mass was obtained by computing
numerically the integral
\begin{equation}
\Mstar=4\pi\int_0^\infty \rhostar(r)r^2\dd r.
\end{equation}
The value of $\Re$ is such that $\Mstarp(\Re)=\Mstar/2$, where
\begin{equation}
\Mstarp(R)=2\pi\int_0^R \Sigmastar(R')R'\,\dd R'
\end{equation}
is the projected stellar mass within $R$ and $\Sigmastar(R)$ is the
total stellar surface density.

Assuming isotropic velocity distribution, the radial component
$\sigmar(r)$ of the velocity dispersion tensor was obtained by solving
the Jeans equation
\begin{equation}
\frac{\dd \left(\rhostar \sigmarsq\right)}{\dd r}=-\rhostar\frac{\dd\Phi}{\dd r},
\end{equation}
where $\Phi(r)$ is the total gravitational potential generated by the
total density distribution $\rhotot=\rhostar+\rhodm$.

For the considered isotropic systems, the line-of-sight velocity
dispersion squared is \citep{Bin82}
\begin{equation}
\sigmalossq(R)={2\over\Sigmastar(R)}\int_R^\infty {\rhostar(r)\sigmarsq r \dd r\over\sqrt{r^2-R^2}}.
\end{equation}
The aperture velocity
dispersion $\sigmaa$ within a projected radius $R$
was determined via
\begin{equation}
\sigmaasq(R)={2\pi\over\Mstarp(R)}\int_0^R \Sigmastar(R')\sigmalossq(R') R'\dd R'.
\end{equation}
The effective stellar velocity dispersion is
$\sigmae\equiv\sigmaa(\Re)$.

\subsection{A reference case}
\label{sec:refcase}

We present here a specific example of the toy model for envelope
accretion described in \Sect\ref{sec:toy}.  We will take this specific
example as a reference case to illustrate quantitatively possible
effects of envelope accretion on the evolution of ETGs in the space of
the parameters $\Mstar$, $\Re$, and $\sigmae$.  For our purposes we do
not need to work in physical units, but we can take as reference mass
and length units, respectively, the stellar mass, $\Mstarini$, and the
effective radius, $\Reini$, of the initial galaxy model. As it is
natural, we adopt as velocity unit
$\vu=\left(G\Mstarini/\Reini\right)^{1/2}$.

The initial galaxy model has stellar density profile given by the
equations of \Sect\ref{sec:stellarcomp} with effective radius
$\Re=\Reini$, $m=4$, and $\Sigmastarzero$ such that
$\Mstar=\Mstarini$. Its DM density profile is given by the equations
of \Sect\ref{sec:darkcomp} with $\rs=10\Reini$ and $\rhos$ such that $\fdme=0.05$, where
\begin{equation}
  \fdme\equiv \frac{\Mdm(\Re)}{\Mstar(\Re)+\Mdm(\Re)}
\label{eq:fdme}
\end{equation}
is the DM fraction within a sphere of radius $\Re$.  Here $\Mstar(r)$
and $\Mdm(r)$ are, respectively, the stellar and DM masses enclosed
within a sphere of radius $r$. For the adopted values of the
parameters, the initial effective velocity dispersion is
$\sigmaeini\simeq 0.296\vu$.
The accreted component has stellar density profile given by the
equations of \Sect\ref{sec:stellarcomp} with effective radius
$\Reaccr=12\Reini$, $m=2$, and $\Sigmastarzero$ such that
$\Mstar=\Mstaraccr=\Mstarini$, and DM density profile given by
\Eq(\ref{eq:rhodmaccr}) with $\Xiaccr=2.5$.
The final galaxy model, which by construction has
$\Mstarfin=\Mstarini+\Mstaraccr=2\Mstarini$, has $\Refin\simeq 3.9\Reini$
and $\sigmaefin\simeq0.293\vu\simeq 0.99\sigmaeini$.

\Fig\ref{fig:rho} shows the stellar and DM density profiles of the
initial galaxy model (dashed curves), the accreted components (dotted
curves), and the final galaxy model (solid curves). The assumed
profiles are somewhat arbitrary, but we note that the relative
distribution of the initial and accreted components is qualitatively
similar to those found in cosmological simulations of massive galaxies
\citep{Ose12,Coo13}.  Though the model shown in \Fig\ref{fig:rho} is
just a specific example of accretion of an extended envelope, it
nicely illustrates that this process can have the effect of producing
an increase in $\Re$ similar to standard mini mergers with $\xi\to 0$,
while keeping $\sigmae$ essentially constant, as it happens in the
case of major mergers.  Modelling, as in \Sect\ref{sec:effect}, the
evolution of $\Re$ and $\sigmae$ with $\Mstar$ with power laws
$\Re\propto\Mstar^\alphaR$ and $\sigmae\propto\Mstar^\alphasigma$, for
the model here considered (and shown in \Fig\ref{fig:rho}) we get
\begin{equation}
\alphaR=\frac{\ln\Refin-\ln\Reini}{\ln\Mstarfin-\ln\Mstarini}\simeq1.96,
\label{eq:alpharenv}
\end{equation}
and
\begin{equation}
  \alphasigma=\frac{\ln\sigmaefin-\ln\sigmaeini}{\ln\Mstarfin-\ln\Mstarini}\simeq-0.01.
  \label{eq:alphasigenv}
\end{equation}

\subsection{Envelope accretion and scaling relations}
\label{sec:env_and_laws}

\begin{figure}
  \includegraphics[width=0.5\textwidth]{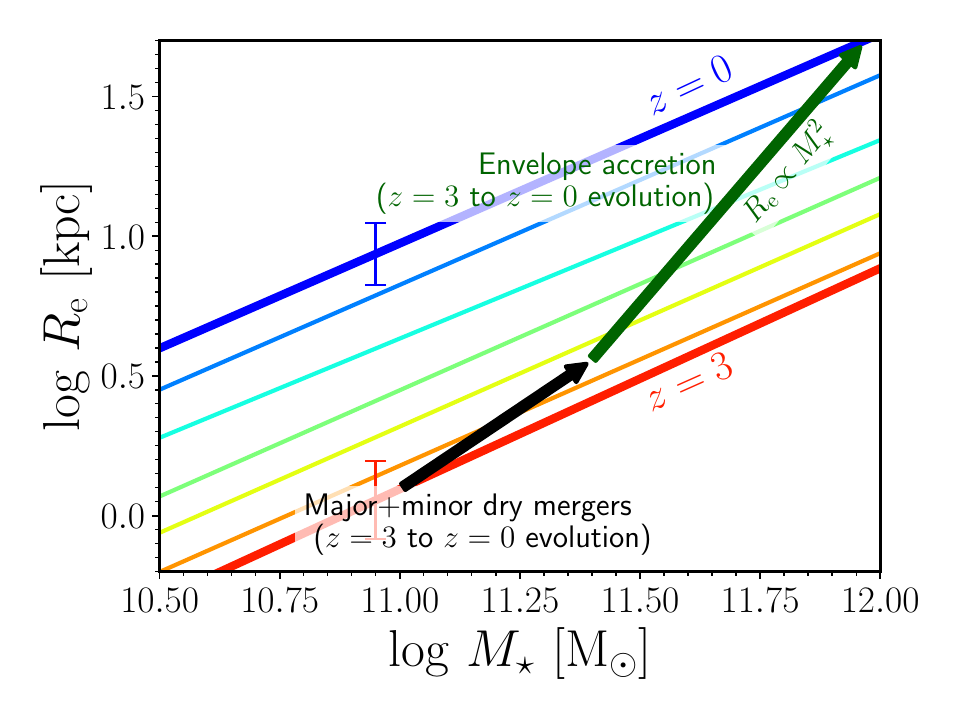}\\ \includegraphics[width=0.5\textwidth]{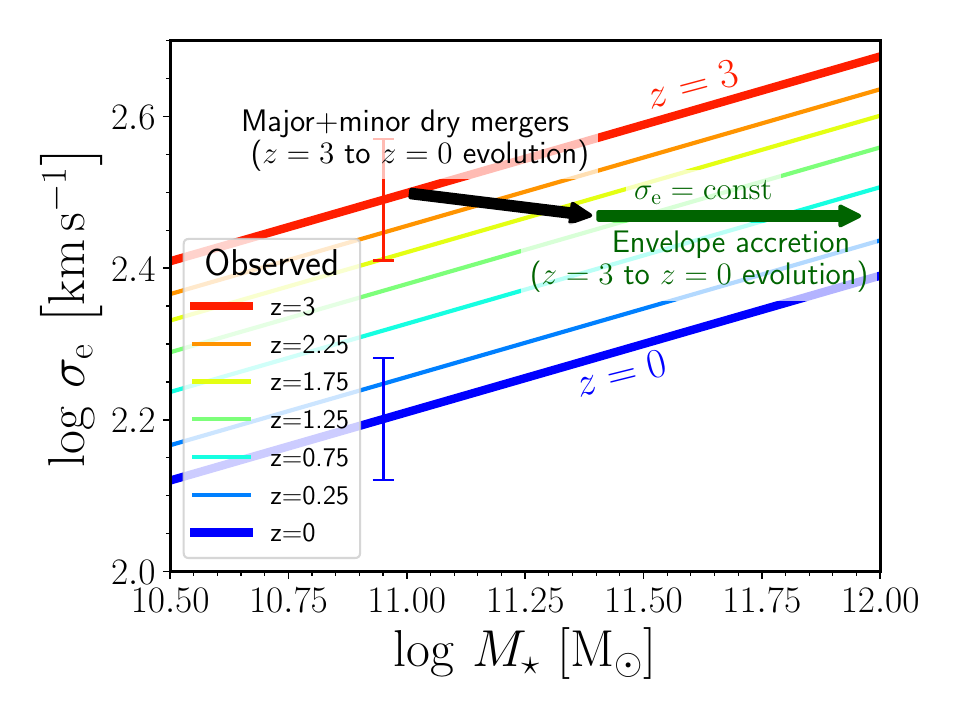}
  \caption{ETG evolution in stellar mass, effective radius, and
    effective velocity dispersion, due to major and minor mergers, and
    envelope accretion. Same as \Fig\ref{fig:scaling_laws_minmaj}, but
    with the addition of the potential contribution from $z=3$ to
    $z=0$ of envelope accretion (green arrows), that is diffuse
    accretion of loosely bound stellar systems such that $\Re$ varies
    with stellar mass as $\Re\propto\Mstar^2$, while $\sigmae$ remains
    constant (see text). While the stellar mass growth attributed to
    major and minor mergers (black arrows) is based on observational
    estimates (\Sect\ref{sec:mergerrate}), the stellar mass growth
    attributed to envelope accretion (green arrows) is chosen ad hoc
    so that the head of the green arrow lies on the observed $z=0$
    $\Re$-$\Mstar$ correlation (upper panel).}
    \label{fig:env_accr}
\end{figure}

\begin{figure*}
  \includegraphics[width=0.5\textwidth]{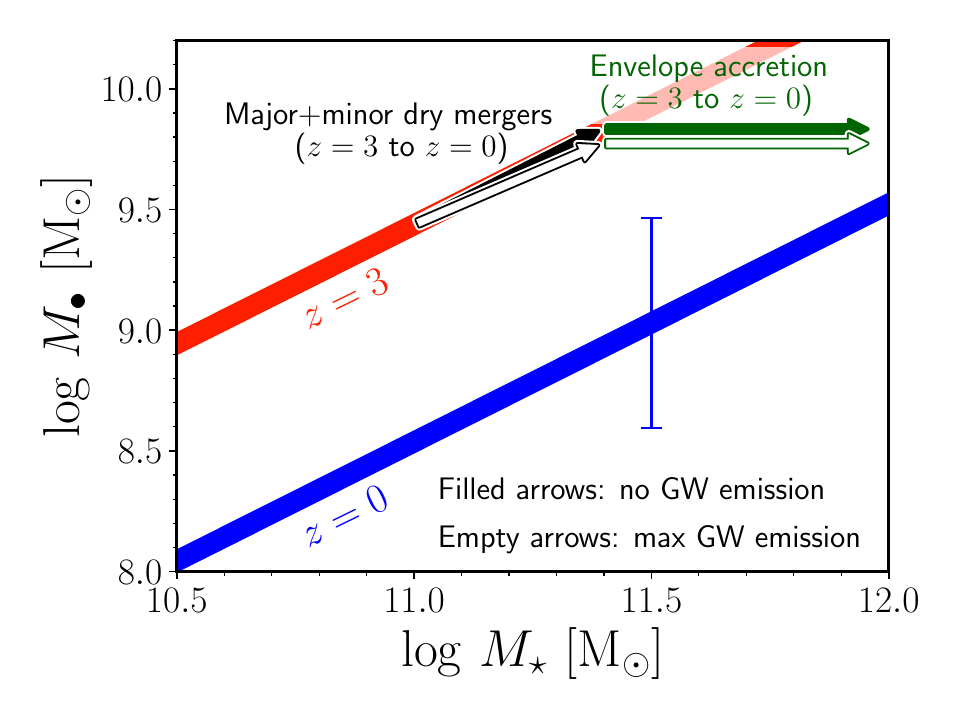}\includegraphics[width=0.5\textwidth]{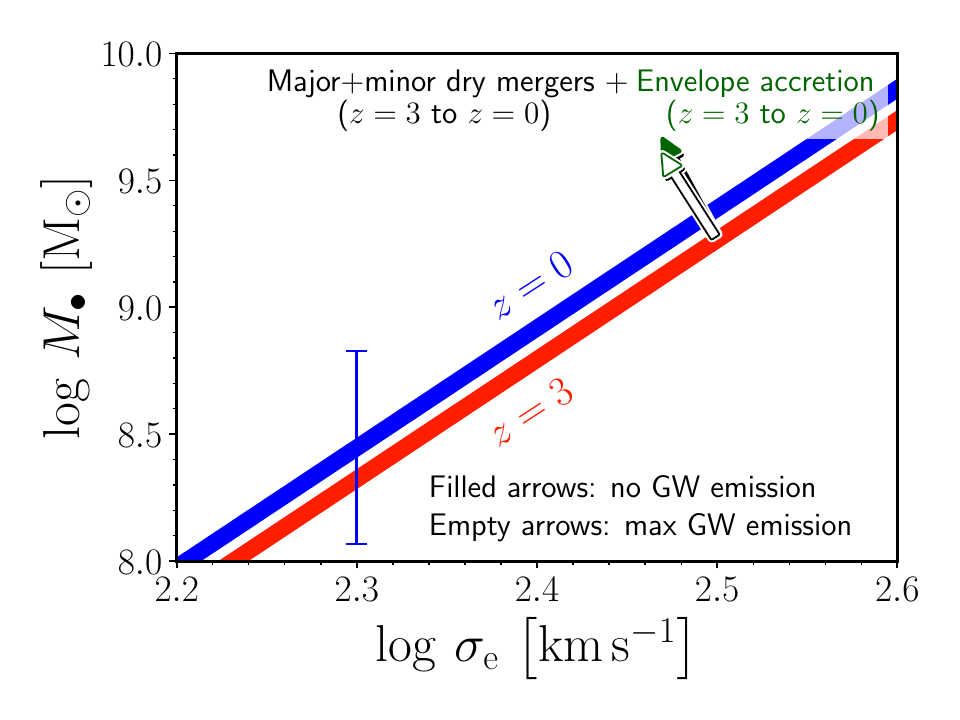}\\
    \caption{ETG evolution in stellar mass, BH mass, and effective
      velocity dispersion, due to major and minor mergers, and envelope
      accretion. Left panel: The blue line is the median relation
      between central BH mass $\Mbh$ and stellar mass $\Mstar$
      measured for observed massive $z=0$ ETGs by \citet{Sag16}, with
      the intrinsic scatter represented by the blue vertical bar. The
      red line is an estimate of the median relationship at $z=3$,
      obtained by assuming that the logarithmic slope is the same as
      at $z=0$ and that the normalization is such that the relic
      galaxy NCG~1277 lies on it (see text). The arrows indicate the
      expected evolution down to $z=0$ of a $z=3$ progenitor due to
      the combination of major and minor mergers, and envelope
      accretion. The filled and empty arrows correspond, respectively,
      to the two extreme cases of no mass loss via GW emission
      ($\Mbh\propto\Mstar$) and of maximal mass loss due to GW
      emission (see text). Right panel: Same as left panel, but for
      the relation between $\Mbh$ and central velocity dispersion
      $\sigmae$. Also in this case the $z=0$ correlation is the median
      obtained by \citet{Sag16}.}
    \label{fig:env_bh}
\end{figure*}

Here we want to study the possible evolution induced by the envelope
accretion in the planes $\Mstar$-$\Re$ and $\Mstar$-$\sigmae$.  Taking
the model presented in \Sect\ref{sec:refcase} as reference to quantify
the effects of envelope accretion on $\Re$ and $\sigmae$, for this
mechanism we assume here $\Re\propto\Mstar^\alphaR$ with $\alphaR=2$
and $\sigmae\propto\Mstar^\alphasigma$ with $\alphasigma=0$ (see
\Eqs\ref{eq:alpharenv} and \ref{eq:alphasigenv}).

In \Fig\ref{fig:env_accr} we report the result of an exercise similar
to that shown in \Fig\ref{fig:standard_mini}, but assuming that,
beside major and minor mergers, the evolution of our reference ETG
from $z=3$ to $z=0$ is due to envelope accretion instead of standard
mini mergers. As done in \Sect\ref{sec:mini} for standard mini
mergers, here the assumed increase in stellar mass attributed to the
envelope is such that our model galaxy lies on the median
$\Re$-$\Mstar$ relation of ETGs at $z=0$.
Comparing the bottom panel
of \Fig\ref{fig:env_accr} with that of \Fig\ref{fig:standard_mini},
the substantial difference in the value of the final $\sigmae$ in the
two models is apparent.  The observationally expected $z=0$ value of
$\sigmae$ lies above that predicted by the standard mini merger model
and below that predicted by the envelope accretion model. We note that
the deviation from the $z=0$ scaling relation of the envelope
accretion prediction is significantly smaller than that of the
standard mini mergers model, though this result should not be taken at
face value, given the highly idealized nature of the considered
models. Nevertheless, this exercise shows very clearly that the
envelope accretion can have an important role in the evolution of ETGs
and that the effect on $\sigmae$ of the accretion of small satellites
(corresponding to mergers with mass ratio lower than $1/10$) could be
quite different from the standard mini merger model, and needs to be
better constrained observationally and theoretically.

Another feature of envelope accretion, which is evident from
\Fig\ref{fig:rho}, is that the inner stellar density distribution is
virtually unchanged during the process. This is particularly
interesting in light of the fact that for massive ETGs (though so far
only over the limited redshift range $0.17<z<0.37$) there is no
evidence of redshift-dependence of the logarithmic slope of the
stellar mass surface density profile within the inner 10 kpc
({\color{blue} Liu et al., submitted}).

{The required fractional mass growth due to envelope accretion
  (see \Fig\ref{fig:env_accr}) is $\fenv\approx 2.6$, which, assuming
  our fiducial values $\fmaj=0.93$ and $\fmin=0.29$
  (\Sect\ref{sec:mergerrate}), corresponds to a ratio between envelope
  accretion growth and major-plus-minor merger  growth
  $\R\equiv\fenv/(\fmaj+\fmin)\simeq 2.1$. This is much higher than
  estimated by \citet{Sue23}, who, observing satellites around massive
  quiescent galaxies in the redshift range $0\leq z \leq 3$, find that
  the stellar mass accreted via mini mergers is $\approx 30\%$ of
  the total accreted stellar mass (corresponding to $\R\approx
  0.43$). However, \citet{Sue23} stress that their estimate must be
  considered a lower limit, because of expected incompleteness of
  their sample.  A rough estimate of the maximum stellar mass
  potentially available for envelope accretion (relative to that for
  major and minor mergers) can be obtained as follows.  Let us model
  the galaxy stellar mass function as a \citet{Sch76} law with
  characteristic mass $\Mstarstar$ and low-mass-end logarithmic slope
  $\alphaSMF$. For a galaxy of stellar mass
  $X\Mstarstar$, the ratio between the stellar mass potentially
  available for mini mergers ($\xi<0.1$) and that potentially
  available for major and minor mergers ($\xi>0.1$) is
  \begin{equation}
\R =   \frac{\int_0^{0.1X}x^{1+\alphaSMF}e^{-x}\dd x}{\int_{0.1X}^X x^{1+\alphaSMF}e^{-x}\dd x},
  \end{equation}
where the integration variable is $x=\Mstar/\Mstarstar$.  For slopes
in the interval $-1.8\lesssim \alphaSMF \lesssim -1.3$ (as observed in
the redshift range $0<z<3$; \citealt{Dav17}) and $X=3$ (i.e. for a
$\Mstar\sim 10^{11}\Msun$ galaxy), we get $0.8\lesssim \R \lesssim
4.5$, so there is potentially room for mini-merger dominated
accretion.  From a theoretical point of view, an independent
indication of the importance of accretion of very small satellites
comes from cosmological simulations. For instance, in DM-only
cosmological simulations, \citet{Fak10} find that the ratio between
mass gained in mergers with DM mass ratio $\xi<0.1$ and that gained in
$\xi>0.1$ mergers is in the range $1.5\lesssim \R \lesssim 2.3$.  }

\section{Evolution of black-hole mass and dark-matter fraction}
\label{sec:bhfdm}

We discuss here the predictions of the considered model involving
envelope accretion for the evolution of properties of the dark
components of quiescent galaxies, namely the mass of the central
supermassive BH and the central DM fraction.

\subsection{Scaling relations involving the mass of the central black hole}
\label{sec:bh}

Little is known observationally about the mass of putative central BHs
in high-redshift quiescent galaxies.
However, useful information in
this respect comes from the study of the relic galaxies, which are
believed to be rare descendants of massive compact high-$z$ ETGs that
have evolved passively since $z\gtrsim 2$. If this is indeed the case,
we can assume that high-$z$ ETGs have the same $\Mbh$ as present-day
relic galaxies of similar $\Mstar$ and $\sigmae$.

Estimates of masses of central BHs in relic galaxies are provided by
\citet{Fer15,Fer17} and \citet{Com23}. They tend to have $\Mbh$ an
order of magnitude higher than non-relic present-day galaxies of
similar stellar mass.  The best studied relic galaxy is NCG~1277, with
$\Mstar\simeq 1.8\times 10^{11}\Msun$ \citep{Com23} and $\sigmae\simeq
385\kms$ \citep{Fer17}, for which \citet{Com23} estimate $\Mbh \simeq
5\times 10^{9}\Msun $.  In the hypothesis that, from the structural
and kinematic point of view, NCG~1277 is representative of $z\approx3$
quiescent galaxies of similar stellar mass, we thus construct
plausible $z=3$ $\Mbh$-$\Mstar$ and $\Mbh$-$\sigmae$ correlations
assuming that they have the same logarithmic slope as the
corresponding $z=0$ correlations, but normalization such that NCG~1277
lies on them.  These $z=3$ correlations are plotted in
\Fig\ref{fig:env_bh} together with the corresponding $z=0$
correlations (with their intrinsic scatter) taken from
\citet[][considering their 'CorePowerE' sample]{Sag16}. These curves
show strong evolution of the $\Mbh$-$\Mstar$ relation, but negligible
evolution of the $\Mbh$-$\sigmae$ relation (within the intrinsic
scatter of the observed $z=0$ relation).  {This behaviour of the
  $\Mbh$-$\Mstar$ and $\Mbh$-$\sigmae$ relations of quiescent galaxies
  is supported by the very recent result of \citet{New25}, who find
  that the massive ($\Mstar\sim 10^{11}\Msun $) gravitationally lensed
  quiescent galaxy MRG-M0138 at $z=1.95$ has $\Mbh/\Mstar$ an order of
  magnitude higher than present-day ETGs, while being consistent with
  the $z=0$ $\Mbh$-$\sigmae$ relation.}

In \Fig\ref{fig:env_bh} we indicate with arrows the predicted
evolution from $z=3$ to $z=0$ of our representative massive ETG in the
$\Mstar$-$\Mbh$ (left panel) and $\sigmae$-$\Mbh$ (right panel)
planes, due to major and minor mergers (black arrows) and to envelope
accretion (green arrows). The $z=3$ BH mass is such that the initial
model galaxy satisfies the $z=3$ $\Mbh$-$\Mstar$ and $\Mbh$-$\sigmae$
correlations.

The model evolution in $\Mstar$ and $\sigmae$, which is the same as in
\Fig\ref{fig:env_accr}, is defined in \Sects\ref{sec:minmajcomparison}
and \ref{sec:env_and_laws}.  The evolution of $\Mbh$ due to major and
minor mergers was modelled assuming that in each merger each galaxy
contains a central BH, that the ratio $\Mbh/\Mstar$ is the same for
both galaxies, and that the two BHs coalesce forming a single BH.  The
mass of the BH remnant is determined by the amount of energy lost via
gravitational waves (GWs), which depends on the properties of the
merger \citep[e.g.][]{Bar12,Mag18}. To bracket a realistic behaviour,
in \Fig\ref{fig:env_bh} we consider two cases: no mass loss via GWs
(filled arrows) and maximal mass loss via GWs (empty arrows). In the
case of no mass loss, the BH mass of the model galaxy grows
proportionally to its stellar mass: $\Mbh\propto\Mstar$. In the case
of maximal mass loss, we heuristically model the growth of the BH as a
power law $\Mbh\propto\Mstar^{\alphabh}$, with
$\alphabh=\ln(1.8)/\ln(2)\simeq 0.85$. With this choice, when the
stellar mass increases by a factor of 2, the BH mass increases by a factor
of 1.8, corresponding to 10\% mass loss via GWs, which is the maximum
expected in a BH-BH merger\footnote{A factor of 2 growth in stellar
mass can be due to combination of mergers with different mass ratios:
for instance a single $\xi=1$ merger or a sequence of ten mergers with
satellites with stellar mass 1/10 of the initial $\Mstar$. It turns
out that the maximum GW-driven BH mass loss is similar in the two
cases, because the maximum fractional mass loss per merger decreases
for decreasing BH mass ratio \citep[][]{Bar12,Mag18}.}, on the basis
of numerical relativity simulations \citep{Bar12}. Given that our
model galaxy grows by a factor $\approx2.5$ via major and minor
mergers, the final BH mass is $\approx 13\%$ lower in the maximal mass
loss case than in the no mass loss case, so uncertainty on the
GW-driven BH mass loss has little effects for our purposes.

The fundamental assumption of the envelope accretion scenario is that
the satellites are disrupted in the galaxy outskirts: if these
satellites possess central BHs, these BHs (due to long
dynamical-friction timescales) are not expected to reach the galactic
nucleus of the main galaxy, but to become wandering BHs. We thus
assume that there is no evolution of $\Mbh$ due to envelope
accretion. As a consequence, the green arrows in \Fig\ref{fig:env_bh}
are horizontal in the $\Mbh$-$\Mstar$ plane (left panel), because only
$\Mstar$ varies, and have zero length in the $\Mbh$-$\sigmae$ plane
(right panel), because neither $\Mbh$ nor $\sigmae$ varies (see also
\Sect\ref{sec:env_and_laws}).

The comparison of the arrows with the $z=3$ ad $z=0$ curves in
\Fig\ref{fig:env_bh} suggests that the combination of (major and
minor) dry mergers and envelope accretion is qualitatively consistent
with the hypothesized evolution of the BH scaling relations.  This
scenario predicts a significantly weaker growth of $\Mbh$ than of
$\Mstar$ and very little evolution in the $\Mbh$-$\sigmae$ plane.
Taken at face value, the results shown in \Fig\ref{fig:env_bh}
indicate that the model slightly overpredicts $\Mbh$ at $z=0$, at given
$\Mstar$ or at given $\sigmae$. However, considering the high
uncertainty of the high-$z$ BH scaling relations, this does not appear as a
significant tension. Moreover, BH mergers might well be less efficient
than assumed in our model, for instance as a consequence of ejection
of BHs due anisotropic GW emission 'kicks' \citep[e.g.][]{Gon07} or
three-body interactions in BH merging hierarchies
\citep[e.g.][]{Hof07}.

\subsection{Dark-matter fraction}
\label{sec:fdm}

High-$z$ quiescent galaxies tend to have lower central DM fraction
than their present-day counterparts \citep[e.g.][]{Men20}. This
finding is also supported by the fact that present-day relic galaxies,
for which the DM can be estimated more robustly than for high-$z$
ETGs, are very DM poor \citep[][]{Com23}.  As it is usual, we define
here the central DM fraction, $\fdme$, as the ratio between the DM mass
and the total mass within a sphere of radius $\Re$ (\Eq\ref{eq:fdme}).
While in high-$z$ (and relic) galaxies $\fdme$ is typically lower than
$\approx 7 \%$ \citep{Men20,Com23}, in normal massive present-day ETGs
$\fdme$ can be higher than $20\%$ \citep{Cap13}.

During the evolution of an individual galaxy, the variation in $\fdme$
can be produced by changes of the amount of DM within a region of
fixed physical size, as well as by changes of $\Re$. In major and
minor dry mergers $\fdme$ increases for the combined effect of the
redistribution of matter and increase in $\Re$. For instance,
\citet{Nip09b} find that the projected DM fraction within $\Re$
increases by up to $\approx 40\%$ in equal-mass mergers and up to a
factor of 2 when the stellar mass doubles with multiple minor mergers.
\citet{Hil13} and \citet{Fri17} find similar results measuring the
intrinsic central DM fraction in binary major and minor merger
simulations.  In the envelope accretion scenario, little DM is added
in the central regions, but $\fdme$ can increase substantially because
of the increase in $\Re$: in other words, after the build-up of the
envelope $\fdme$ is measured over a volume including more DM dominated
regions. This is apparent from \Fig\ref{fig:rho}: in the reference
model shown there, the central DM fraction raises from $\fdme=0.05$ in
the initial configuration to $\fdme\simeq 0.42$ in the final
configuration.

Though a quantitative comparison between the observed and predicted
evolution of $\fdme$ appears premature, qualitatively, major and minor dry
mergers, as well as envelope accretion, are promising processes to
explain the increase in $\fdme$ with cosmic time.

\section{Conclusions}
\label{sec:concl}

We have studied the structural and kinematic evolution of massive
quiescent galaxies from $z\approx 3$ to $z\approx 0$ using
state-of-the-art measurements of the evolution of the scaling
relations and of the merger rates. Our main conclusions are the following.
\begin{itemize}
\item Major and minor mergers (with mass ratio $\xi>1/10$) at
  observationally motivated rates produce, over the redshift range
  $0\lesssim z\lesssim 3$, size and velocity dispersion evolution
  significantly weaker than that measured for ETGs, accounting only
  for $\approx 6\% $ of the evolution in $\Re$ and $\approx 40\%$
  of the evolution in $\sigmae$.
\item The poorly constrained contribution of mini mergers (with mass
  ratios $\xi<1/10$) can compensate the weak evolution induced by
  higher-$\xi$ mergers, but only if these small satellites are so
  diffuse that they are disrupted in the galaxy outskirts, leaving the
  more central regions of the galaxy almost untouched.  In this
  envelope accretion scenario, size grows fast with stellar mass
  (approximately as $\Re\propto\Mstar^2$), while $\sigmae$ remains
  essentially constant.
\item A model in which quiescent galaxies grow via envelope accretion,
  in addition to major and minor mergers, predicts, at given stellar
  mass, higher central BH mass at higher redshift.  This trend is
  consistent with the (though not so stringent) currently available
  observational constraints on the evolution of the $\Mbh$-$\Mstar$
  and $\Mbh$-$\sigmae$ relations of quiescent galaxies.
\item Envelope accretion and dry mergers can also explain, at least
  qualitatively, the finding that the central DM fraction of quiescent
  galaxies increases with cosmic time.
\end{itemize}  

The starting point of our investigation was highlighting
quantitatively the fact that major and minor dry mergers at realistic
rate are grossly unable to explain the evolution of the $\Re$-$\Mstar$
and the $\sigmae$-$\Mstar$ relation.  We have presented an idealized
model of envelope accretion that shows that mini mergers can help fill
the gap in the evolution of $\Re$ and $\sigmae$, if the accreted
satellites deposit the vast majority of their dark and luminous matter
far from the galactic centre.

{A potential problem is that, in order to explain the observed
  size evolution, our model requires that the stellar mass growth via
  envelope accretion is twice the growth via major and minor mergers,
  while there is no observational evidence, so far, of such a
  predominance of very small satellite accretion (see
  \Sect\ref{sec:env_and_laws} for a discussion).  However, a smaller
  contribution from envelope accretion would be required if the
  merger-driven size increase were stronger (see
  \Sect\ref{sec:validity}) or the evolution of the $\Re$-$\Mstar$
  relation were weaker (see \Sect\ref{sec:remstarz}) than considered
  in this paper.  Moreover, our model does not include the well-known
  effect known as progenitor bias \citep[e.g.][]{Fra08}, which, though
  expected to have a minor role for the very high-mass galaxies here
  considered \citep[][see also \citealt{Cla25}]{Fag16}, would go in
  the direction of reducing the required mass and size growth due to
  envelope accretion.

Our treatment of envelope accretion is a proof of concept that can be
taken as a starting point for theoretical studies aimed at modelling
more quantitatively the build-up of envelopes with numerical
simulations. From the observational point of view it would be
beneficial to improve our knowledge of the redshift-dependence of the
$\sigmae$-$\Mstar$ relation and of $\fdme$ ({\color{blue} Cannarozzo
  et al., in prep.}), of the scaling relations involving BH masses (in
the spirit of \citealt{New25} and \citealt{Tan25}), and also,
following \citet{Sue23}, of the rate and properties of mini mergers
through the observation of very low-mass satellites around massive
quiescent galaxies.  }

\begin{acknowledgements} 
I am grateful to an anonymous referee for comments that helped improve
the paper.  The research activities described in this paper have been
co-funded by the European Union – NextGenerationEU within PRIN 2022
project n.20229YBSAN - Globular clusters in cosmological simulations
and in lensed fields: from their birth to the present epoch.
\end{acknowledgements}

\bibliographystyle{aa}
\bibliography{biblio_envelo.bib}

\end{document}